\documentclass[a4paper,11pt]{article}
\usepackage{pos}

\def\mC{\ensuremath{\mathcal{C}}}
\def\mD{\ensuremath{\mathcal{D}}}
\def\mF{\ensuremath{\mathcal{F}}}

\def\mP{\ensuremath{\mathcal{P}}}

\def\mS{\ensuremath{\mathcal{S}}}
\def\mT{\ensuremath{\mathcal{T}}}

\title{Functional methods for hadron spectroscopy}
\author*[a]{Gernot Eichmann}
\author[b,c]{Christian S. Fischer}
\author[a,b,c]{Joshua Hoffer}

\affiliation[a]{University of Graz, NAWI Graz,\\
  Universitätsplatz 5, 8010 Graz, Austria}

\affiliation[b]{Institut für Theoretische Physik, Justus-Liebig-Universität Gießen\\
Heinrich-Buff-Ring 16, 35392 Gießen, Germany}

\affiliation[c]{Helmholtz Forschungsakademie Hessen für FAIR (HFHF),\\
GSI Helmholtzzentrum für Schwerionenforschung, Campus Gießen\\
Heinrich-Buff-Ring 16, 35392 Gießen, Germany}

\emailAdd{gernot.eichmann@uni-graz.at}
\emailAdd{christian.fischer@theo.physik.uni-giessen.de}
\emailAdd{joshua.hoffer@uni-graz.at}

\abstract{We summarize recent results for four-quark states with hidden and open flavor
obtained from a four-quark Bethe-Salpeter/Faddeev-Yakubowsky equation. 
The approach dynamically predicts the leading
internal two-body clusters such as meson-meson, hadroquarkonium or diquark-antidiquark components.
For hidden-flavor states, the meson-meson or hadroquarkonium configurations are dominant,
while the diquark contributions are small. For open-flavor states, the situation is different and
depends on the quark content: The $T_{bb}^-$ is a mixture of various components, where the diquark-antidiquark
components are dominant, whereas the $T_{cc}^+$ is practically entirely dominated by $DD^\ast$ due to its proximity
to the threshold. We also provide details on the analytic continuations and extrapolations to extract states above thresholds.}

\FullConference{The XVIth Quark Confinement and the Hadron Spectrum Conference (QCHSC24)\\
 19-24 August, 2024\\
 Cairns Convention Centre, Cairns, Queensland, Australia\\}

\begin{document}
\maketitle

\begin{figure}[t!]
	\centerline{ %
		\includegraphics[width=0.8\textwidth]{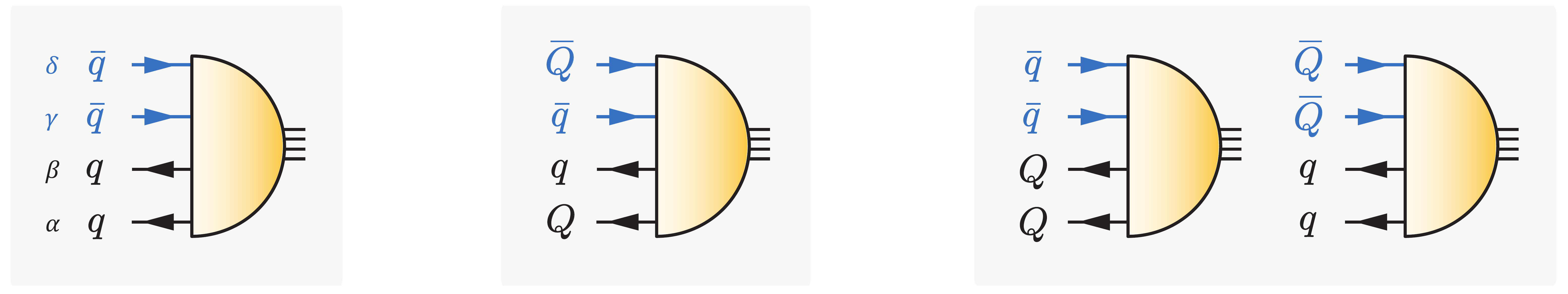}}
	\caption{Different configurations for four-quark states: four equal quarks ($qq \bar{q}\bar{q}$), hidden flavor ($Q\bar{Q}q\bar{q}$)
             and open flavor ($QQ\bar{q}\bar{q}$, $qq\bar{Q}\bar{Q}$). }
	\label{fig:3confs}
\end{figure}

\section{Introduction}

Over the past two decades much experimental progress has been made in mapping out the spectrum of exotic hadrons,
which do not fit into the conventional picture of mesons as quark-antiquark ($q\bar{q}$) states and baryons made of three valence quarks ($qqq$).
There is evidence for four-quark states or tetraquarks ($q\bar{q}q\bar{q}$) and pentaquarks ($qqqq\bar{q}$),
and searches for hybrid mesons with extra gluonic components and glueballs, which are made of gluons only, are ongoing~\cite{Chen:2016qju,Lebed:2016hpi,Esposito:2016noz,Ali:2017jda,Olsen:2017bmm,Guo:2017jvc,Liu:2019zoy,Brambilla:2019esw}.

States involving heavy quark flavors are of particular interest. 
Unlike light mesons, where relativity and chiral symmetry play a significant role, 
the heavy-quark sector offers a much cleaner environment for studying exotic states. 
The success of the non-relativistic quark model provides a benchmark for distinguishing conventional mesons from exotic ones, 
based on deviations from quark-model predictions.
The recent abundance of data in the charm sector has firmly established the existence of four-quark states. 
Notably, many of these states lie near meson-meson thresholds which suggests a molecular nature~\cite{Guo:2017jvc}.

In the following we focus on two types of exotic mesons, see Fig.~\ref{fig:3confs}. The first are \textit{hidden-flavor states} of the form  $Q\bar{Q}q\bar{q}$, where $Q = c, b$ is a heavy quark
       and $q = u, d, s$ is a light quark. If these states carry isospin $I=0$, they can still mix with conventional $Q\bar{Q}$ mesons;
       otherwise, they are manifestly exotic.
       The second type are \textit{open-flavor states} of the form $QQ\bar{q}\bar{q}$ and $qq\bar{Q}\bar{Q}$, which are manifestly exotic as they cannot be constructed from a simple quark-antiquark pair.
       
       Concerning hidden-flavor states, the most prominent example is the $\chi_{c1}(3872)$, which has quantum numbers $I(J^{PC}) = 0(1^{++})$
       and was first reported by Belle in 2003~\cite{Belle:2003nnu}. 
       Several of its properties  do not align with a conventional charmonium interpretation, including
       a mass nearly identical with the $D^0 \bar{D}^{\ast 0}$ threshold,
        a narrow width ($<1.2$ MeV) which is significantly smaller than potential-model predictions for the first excited
        $c\bar{c}$ state, and unusual decay patterns that deviate from standard charmonium expectations.
        Other exotic candidates, such as the $\psi(4230)$ in the $1^{--}$ vector channel,
        are considered exotic mainly due to the overpopulation of this channel compared to well-established  $c\bar{c}$ states.
        In contrast, the $Z_c(3900)$ and $Z_c(4430)$, both with $J^{PC} = 1^{+-}$, are  manifestly exotic because they carry electric charge. 
        Their minimal quark content must be $c\bar{c}u\bar{d}$, which rules out a simple $c\bar{c}$ interpretation.
        
      For open-flavor states with heavy quarks, only one state has been experimentally confirmed so far:
      the $T_{cc}(3875)^+$ with quark content $cc\bar{u}\bar{d}$,
      which is very close to the $D^0 D^{\ast +}$ threshold~\cite{LHCb:2021vvq}.
      If the heavy-quark masses are sufficiently large while the light-quark masses remain small,
      such states are expected to form  bound systems~\cite{Francis:2016hui,Eichten:2017ffp}.
      A notable example is the theoretically predicted 
      $T_{bb}^-$ made of $bb\bar{u}\bar{d}$, which is supported by many calculations~\cite{Hudspith2023,Alexandrou2024,Junnarkar2019,Leskovec2019,Aoki2023,
      Park2019,Noh2021,Maiani2019,Braaten2021}.  
      
      Additionally, LHCb has observed states with one heavy quark ($Q\bar{q}q\bar{q}$) and 
      even fully heavy tetraquark states ($Q\bar{Q}Q\bar{Q}$),
      which further expands the growing list of heavy exotic hadrons~\cite{LHCb-FIGURE-2021-001-report}.
      
      An important question concerns the internal structure of such states.
      For hidden-flavor states, the quarks and antiquarks may arrange themselves into different configurations.
      They could form tightly bound diquark-antidiquark pairs $(Qq)(\bar{Q}\bar{q})$ interacting via colored forces~\cite{Esposito:2016noz}. 
      Alternatively, the heavy quark
      and antiquark could cluster together into a compact core surrounded by a light $q\bar{q}$ pair, forming a $(Q\bar{Q})(q\bar{q})$ configuration,
      which is the hadro-quarkonium picture~\cite{Voloshin:2007dx}. Or the state could cluster into two heavy-light meson components $(Q\bar{q})(q\bar{Q})$,
      which is particularly relevant for states close to meson-meson thresholds and then corresponds to a meson-molecule picture~\cite{Guo:2017jvc}.
      For open-flavor states the possible clusters are different due to their quark content: 
      The heavy quarks could form a diquark pair bound to an antidiquark, $(QQ)(\bar{q}\bar{q})$,
      or the state could consist of two degenerate meson-meson configurations $(Q\bar{q})(Q\bar{q})$. 
      Which of these configurations is realized is a dynamical question. In general 
      these possibilities are also not mutually exclusive: In principle every experimental state may be a superposition
      of components with a different structure, where the leading component may differ on a case-by-case basis.
      If the quantum numbers allow it, also substantial mixing effects with `ordinary' quark-antiquark mesons may occur.
      
In order to interpret exotic hadrons like glueballs, hybrids, four- and five-quark states, it is of utmost importance
to develop theoretical approaches in QCD that are able to cover all these states in a single framework and shed
light on the dynamical mechanism responsible for their internal structure. To this end, functional methods, in
particular the combination of Dyson-Schwinger equations (DSEs), Bethe-Salpeter equations (BSEs) and Faddeev
equations, have matured substantially over the last years~\cite{Roberts:1994dr,Alkofer:2000wg,Bashir:2012fs,Eichmann:2016yit,Huber:2018ned,Fischer:2018sdj}. 
They provide
insight in the nonperturbative properties of QCD from dynamical mass generation to the spectrum and interactions
of hadrons. This establishes
a promising avenue for spectrum and structure calculations of exotic hadrons that can complement lattice QCD
calculations, amplitude analyses and phenomenological approaches.

\section{Four-body equation}

One way to calculate bound states or resonances of quarks and gluons is to solve covariant bound-state equations or BSEs.
These are the quantum-field theoretical analogs of the Schrödinger equation in quantum mechanics,
which allow one to calculate the spectrum and (Bethe-Salpeter) wave functions of hadrons.
Originally formulated for two-body systems~\cite{Salpeter:1951sz,Llewellyn-Smith:1969bcu,Nakanishi:1969ph}, 
these equations can be extended to describe three-, four-, or even multi-particle states, 
in which case they are also known as Faddeev and Faddeev-Yakubowski equations~\cite{Faddeev:1960su,Yakubovsky:1966ue}.

\begin{figure*}[t]
                    \centering
                    \includegraphics[width=1.0\textwidth]{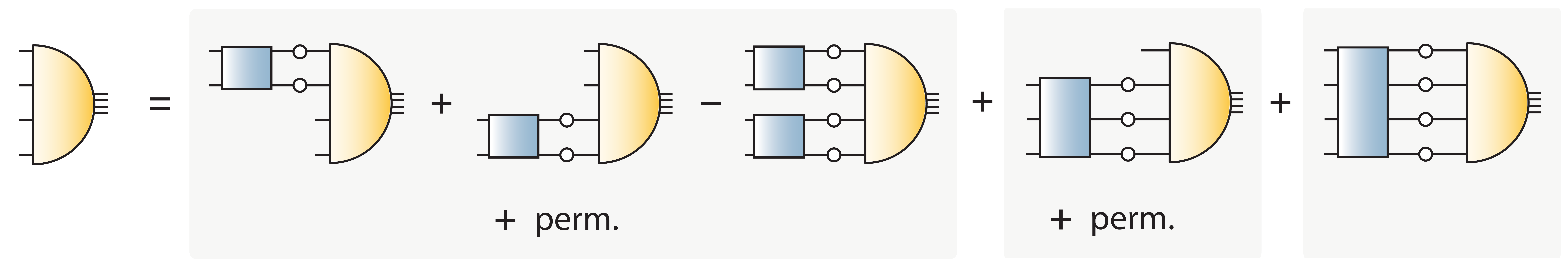}
                    \caption{Four-quark Bethe-Salpeter equation with two-, three- and four-body kernels.
                    The lines with circles are dressed quark propagators.
                     }\label{fig:4q-bse}
                     \vspace{2mm}
            \end{figure*}

Fig.~\ref{fig:4q-bse} shows the BSE (or Faddeev-Yakubowski equation)  for a four-quark system.
It is a homogeneous integral equation for the four-quark amplitude. The kernel of the equation 
consists of two-, three- and four-particle  interactions.
Restricting ourselves to  two-particle  interactions, we are left with   the first three terms on the r.h.s.,
where the subtraction terms are necessary to prevent overcounting~\cite{Huang:1974cd,Khvedelidze:1991qb,Heupel:2012ua}.
If we label the quarks in the $qq\bar{q}\bar{q}$ amplitude by 1, 2 and the antiquarks by 3, 4, 
then the kernel allows for three possible permutations: (12)(34) as shown in the figure, which
describes interactions in the $qq$ and $\bar{q}\bar{q}$ (`diquark') channels, plus (13)(24) and (14)(23) 
which describe interactions in the $q\bar{q}$ (`meson') channels. 
Clearly, by iterating the equation the two-quark kernels generate further (reducible) three- and four-quark interactions.

The four-body equation in Fig.~\ref{fig:4q-bse} is an exact equation.
It can be derived from the scattering equation for the $qq\bar{q}\bar{q} \to qq\bar{q}\bar{q}$
eight-point function, whose poles describe hadron bound states and resonances.
At a given pole, the residue of the scattering equation is the homogeneous four-body equation shown in Fig.~\ref{fig:4q-bse}.
The unknowns are then the two-, three- and four-body kernels and the dressed quark propagator.
The benefit of the approach is that these same unknowns also appear in the equations for $q\bar{q}$ and $qqq$ systems.
In particular, if we drop the three- and four-body irreducible kernels in  Fig.~\ref{fig:4q-bse},
the remaining inputs are the two-body kernels and the quark propagator.
Thus, once these expressions are known one can calculate the properties of
mesons, baryons and four-quark states without any further input.

The quantities that enter in these expressions are QCD's $n$-point correlation functions:
the two-point functions like the quark  and gluon propagators,
three-point functions like the quark-gluon and three-gluon vertex, etc.
These satisfy DSEs~\cite{Roberts:1994dr,Alkofer:2000wg,Bashir:2012fs,Eichmann:2016yit,Huber:2018ned,Fischer:2018sdj}, 
which are the quantum equations of motion
of a quantum field theory. These form an infinite set of coupled integral equations
which relate QCD's $n$-point correlation functions with each other.
Alternatively, the $n$-point functions can be computed using the functional renormalization group, 
which leads to coupled integral-differential equations~\cite{Berges:2000ew,Pawlowski:2005xe,Dupuis:2020fhh}.

\begin{figure*}[t]
                    \centering
                    \includegraphics[width=1.0\textwidth]{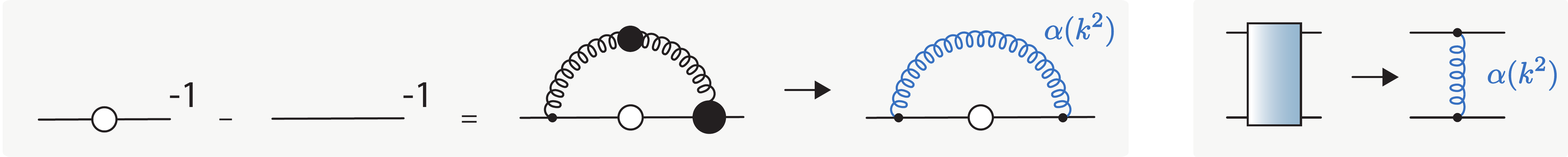}
                    \caption{Rainbow-ladder truncation for the quark DSE and quark-(anti-)quark kernel.
                     }\label{fig:rl}
            \end{figure*}

Over the past decades, a broad range of hadron properties has been successfully calculated in this approach:
masses and decay constants of light, heavy and heavy-light mesons; the spectra of light and heavy baryons,
   electromagnetic elastic and transition form factors of mesons and baryons, parton distribution functions 
   and scattering amplitudes, see e.g.~\cite{Maris:2005tt,Bashir:2012fs,Eichmann:2016yit,Barabanov:2020jvn}
   and references therein.
  In many  hadron physics applications, one  bypasses the task of solving coupled DSEs
   by making a rainbow-ladder ansatz for the two-body ($qq$ and $q\bar{q}$) kernels
   with an effective quark-gluon interaction, such that only the quark DSE needs to be solved explicitly (see Fig.~\ref{fig:rl}).
   This preserves chiral symmetry and thereby guarantees the chiral properties of the pion, i.e.,
   the pion is automatically massless in the chiral limit. Correspondingly, the quark DSE
    shows dynamical chiral symmetry breaking by generating a large, dynamical quark mass function at low momenta,
   which is the relevant mass scale for hadrons (see~\cite{Eichmann:2025wgs} for a pedagogical discussion).

   In parallel, efforts are underway to perform calculations beyond rainbow-ladder 
   which incorporate corrections from dynamical gluons, vertex dressings, and higher-order interaction effects~\cite{Chang:2009zb,Eichmann:2016yit,Huber:2020keu,Eichmann:2021zuv,Ferreira:2023fva}. 
   However, due to their significantly higher computational cost,  
   their widespread application in hadron physics is only beginning to take shape 
   and is expected to advance in the near future.
   A recent example is the glueball sector in Yang-Mills theory, where a 
    consistent solution of the underlying DSEs yields a glueball spectrum
    in agreement with lattice results~\cite{Huber:2021yfy}.

   In our discussion of four-quark states we will  follow the rainbow-ladder approach.   
   As shown in Fig.~\ref{fig:rl}, the quark DSE turns into a standalone equation which only depends on an effective quark-gluon interaction $\alpha(k^2)$
   that absorbs all further information. The BSEs for hadrons are then solved using a two-body kernel
   with the same ansatz for $\alpha(k^2)$, which is meant to absorb all further kernel contributions.
     A popular ansatz is the 
  Maris-Tandy model~\cite{Maris:1999nt}, $\alpha(k^2) = \pi\,\eta^7 x^2\,e^{-\eta^2 x} + \alpha_\text{UV}(x)$ with $x=k^2/\Lambda^2$.
  Here the  second term $\alpha_\text{UV}(x)$ ensures the correct perturbative behavior at large momenta but is otherwise not  important.
  The relevant term is the first one, which effectively only depends  on the scale $\Lambda = 0.72$ GeV that is chosen to reproduce the 
   pion decay constant $f_\pi \approx 92$ MeV. If one varies the shape parameter $\eta$,
  the shape of the interaction may look very different, but many observables calculated with it are insensitive to this change.
  For this reason, also other forms of $\alpha(k^2)$ proposed in the literature lead to very similar results for observables~\cite{Qin:2011dd}.

\section{Two-body poles}

Returning to the four-body equation in Fig.~\ref{fig:4q-bse}, the question remains how the equation is solved in practice.
The four-body amplitude has the form
\begin{equation}\label{4q-amp}
   \Gamma^{(\mu)}(p,q,k,P)_{\alpha\beta\gamma\delta} = \sum_{i=1}^N f_i(\Omega)\,\tau_i^{(\mu)}(p,q,k,P)_{\alpha\beta\gamma\delta} \otimes \text{Color}\otimes\text{Flavor}\,.
\end{equation}
Here, $p$, $q$ and $k$ are the  relative momenta that can be formed from the four (anti-)quark momenta, and $P$ is the total momentum.
The amplitude carries four Dirac indices, one for each (anti-)quark leg, and for $J=1$ states there is an additional Lorentz index $\mu$.
This leads to $N= 256$ Dirac tensors for $J=0$ states and $N=768$ Dirac-Lorentz tensors for $J=1$~\cite{Eichmann:2015cra,Wallbott:2019dng}.
Even worse, the dressing functions $f_i(\Omega)$ depend on ten Lorentz invariants
\begin{equation}\label{kin-var}
   \Omega = \{ p^2, \, q^2, \, k^2,\, p\cdot q, \, p\cdot k, \, q\cdot k, \, p\cdot P, \, q\cdot P, \, k\cdot P, \, P^2= -M^2 \}\,,
\end{equation}
where the total momentum $P$ is onshell and $M$ is the mass of the four-quark state.
In practice, the BSE turns into an eigenvalue equation for a kernel matrix of the size $\sim 10^{13}$ depending on the number of grid points. Clearly,
this is impossible to solve even with the best computers available today, and one needs to make approximations on the structure of the amplitude.

To this end one can ask: What are the \textit{relevant} tensors $\tau_i^{(\mu)}$ that capture the important dynamics?
Presumably not all of them are equally important. From two- and three-body systems it is known that the momentum-independent tensors
are the most important ones, while those  depending on higher momentum powers
are increasingly suppressed. In a partial-wave expansion, where the tensors are arranged into eigenstates
of the quark orbital angular momentum in the hadron's rest frame,
the momentum-independent tensors are  the `$s$ waves' with orbital angular momentum $L=0$.
%which contribute the leading effects. 
This leads to 32 $s$-wave tensors for $J=0$ states and 48 for $J=1$ states~\cite{Eichmann:2015cra,Wallbott:2019dng}.

Similarly, not all kinematic variables in Eq.~\eqref{kin-var} will be equally important,
but a priori it is difficult to judge which ones can be safely neglected.
If we arbitrarily dropped some of those, this would also break
the permutation symmetries of the four-quark system, namely the Pauli antisymmetry when exchanging
two identical quarks and the charge-conjugation symmetry when exchanging quark and antiquark.
To this end, it is useful to arrange these ten variables in multiplets which transform under irreducible representations of the permutation group $S_4$~\cite{Eichmann:2015nra,Eichmann:2015cra}.
This yields a singlet $\mS_0 \sim p_1^2 + p_2^2 + p_3^2 + p_4^2$,
which carries the scale, as well as a  doublet $\mD$ and two triplets $\mT_1$, $\mT_2$.
Together with the singlet $P^2 = -M^2$, one has thus 1 + 2 + 3 + 3 + 1 = 10 variables.

\begin{figure*}[t]
                    \centering
                    \includegraphics[width=0.75\textwidth]{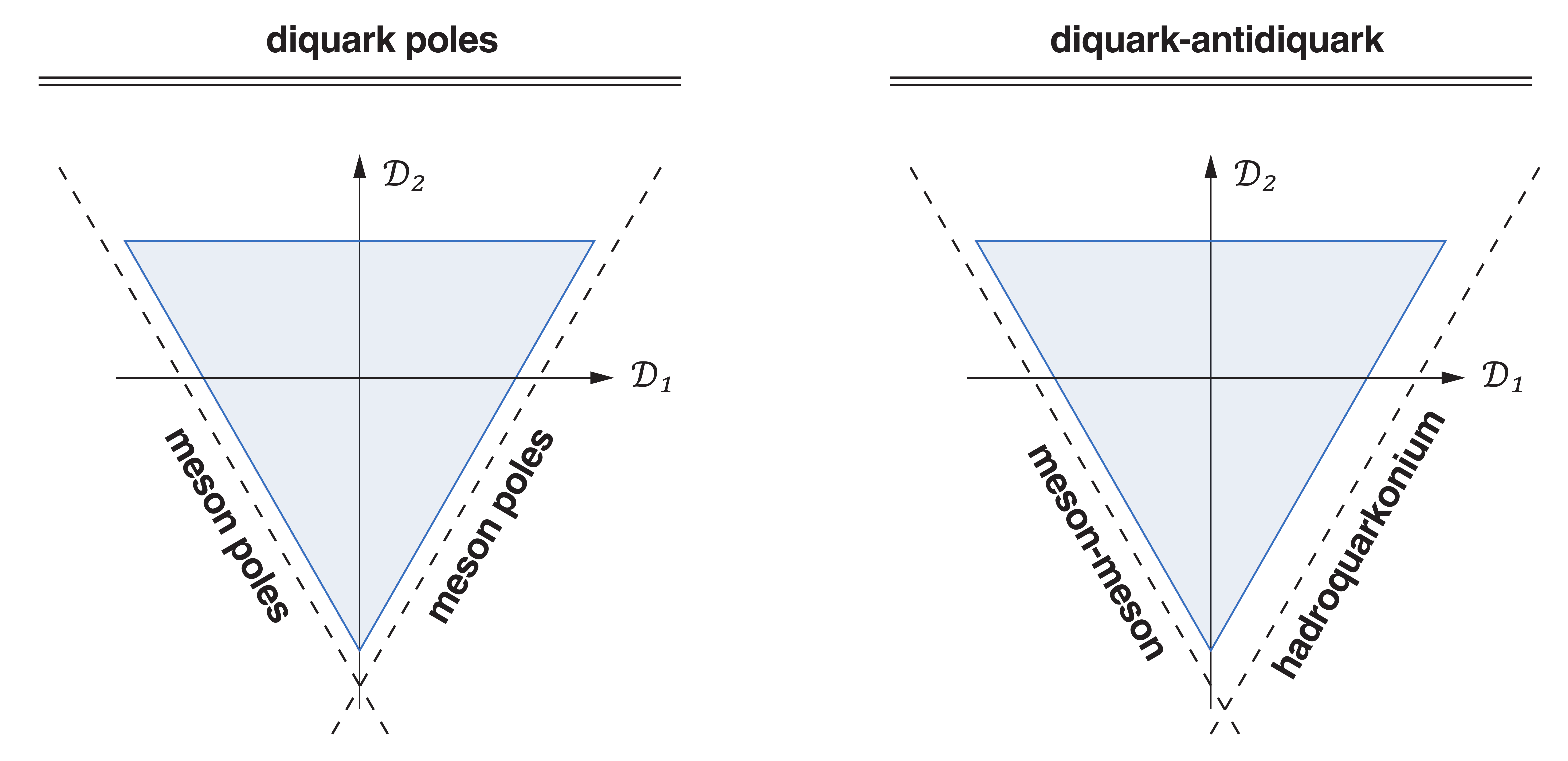}
                    \caption{Mandelstam plane in the doublet variables $\mD_1$ and $\mD_2$. Left: Four light quarks 
                              dynamically create light pions, whose poles are close to the integration
                             region (colored triangle), and heavier diquarks whose poles are farther away.
                             Right: For heavy-light four-quark states with hidden flavor, 
                             the possible two-body clusters correspond to diquark-antidiquark, meson molecule and hadroquarkonium configurations.
                     }\label{fig:4q-poles}
            \end{figure*}

The solution of the four-body equation for light quarks shows that the singlet $\mS_0$ and doublet $\mD$ 
are the most important variables, while one triplet has only modest effects and the other  is almost irrelevant~\cite{Eichmann:2015cra}.
This can be understood on physical grounds, because the doublet contains the two-body poles which are dynamically generated in the equation.
In the solution of the BSE, the four-body amplitude dynamically develops meson and diquark poles  which show up in the two doublet variables.
This is illustrated in the left panel of Fig.~\ref{fig:4q-poles}: for the $\sigma/f_0(500)$ meson made of four light quarks,
the resulting pion poles are close to the integration region (the colored triangle) and therefore dominate the system,
while the diquark poles are far away and much less important. The four-quark BSE then  
reproduces the mass pattern for the light scalar mesons~\cite{Eichmann:2015cra}. This leads to an interpretation of the $\sigma$ meson
as being dominated by $\pi\pi$ components, which is in line with many studies in the literature~\cite{Pelaez:2015qba}. 

For heavy-light four-quark states, the analogous situation is sketched in the right panel of Fig.~\ref{fig:4q-poles}:
In this case, the three sides of the triangle correspond to diquark-antidiquark, meson molecule and hadroquarkonium.
These are just the three possibilities for internal two-body clustering mentioned in the introduction.
In this way, the four-body equation dynamically decides which of these clusters is more important without making prior assumptions.
Given that the meson poles are typically closer  to the integration region and the diquark poles farther away, 
this hints towards meson-meson components as the dominant internal structures, 
although the interplay of the tensors in Eq.~\eqref{4q-amp} due to the quark-gluon dynamics in the BSE can further complicate this picture.
 
To make the four-body equation feasible for numerical calculations, we should therefore keep at least $\mS_0$ and $\mD$
as dynamical variables. In practice we go a step further and keep only those `physical' tensors in Eq.~\eqref{4q-amp}
whose Dirac-color-flavor structures match the relevant internal meson and diquark correlations that can contribute to a given state.
For example, the leading components for the $\chi_{c1}(3872)$ with $I(J^{PC}) = 0(1^{++})$ correspond to $DD^\ast$, $J/\psi \omega$ and
scalar-axialvector diquark configurations, whose leading Dirac-color-flavor tensors are given by~\cite{Wallbott:2020jzh}
\begin{equation}
\begin{split}
   (\tau_0^\mu)_{\alpha\beta,\gamma\delta} &= (\gamma^5_{\alpha\gamma}\,\gamma^\mu_{\beta\delta} - \gamma^\mu_{\alpha\gamma}\,\gamma^5_{\beta\delta})\,\mC_{11}\,\mF_0\,, \\
   (\tau_1^\mu)_{\alpha\beta,\gamma\delta} &= \varepsilon^{\mu\nu\rho\sigma}\,\hat{P}^\nu\,\gamma^\rho_{\alpha\delta}\,\gamma^\sigma_{\beta\gamma}\,\mC'_{11}\,\mF_0\,, \\
   (\tau_2^\mu)_{\alpha\beta,\gamma\delta} &= \left( (\gamma_5 C)_{\alpha\beta}\,(C^T \gamma^\mu)_{\gamma\delta} - (\gamma^\mu C)_{\alpha\beta}\,(C^T \gamma_5)_{\gamma\delta}\right)\mC_{\bar{3}3}\,\mF_0\,. \\
\end{split}
\end{equation}
The Dirac indices are ordered as in Fig.~\ref{fig:3confs}, with $\alpha$, $\beta$ for the quarks and $\gamma$, $\delta$ for the antiquarks.
$C = \gamma^4 \gamma^2$ is the charge-conjugation matrix, and $\hat{P}^\nu$ is the normalized total momentum. 
The flavor wave function in the isospin $I=0$ channel is $\mF_0 = c(u\bar{u}+d\bar{d})\bar{c}$, and
$\mC_{11}$, $\mC'_{11}$ and $\mC_{\bar{3}3}$ are the  color tensors
for the meson-meson ($\mathbf{1} \otimes \mathbf{1}$) and diquark-antidiquark ($\overline{\mathbf{3}}\otimes \mathbf{3}$) configurations,
of which only two are independent:
\begin{equation}
   \mC_{11} = \frac{1}{3}\,\delta_{AC}\,\delta_{BD}\,, \qquad
   \mC'_{11} = \frac{1}{3}\,\delta_{AD}\,\delta_{BC}\,, \qquad
   \mC_{\bar{3}3} = -\frac{\sqrt{3}}{2}\,(\mC_{11} - \mC'_{11})\,.
\end{equation}

Furthermore, because we already know that the four-quark equation  dynamically generates meson and diquark poles
in the doublet variables, we absorb these poles directly in the tensors by writing
\begin{equation}
\begin{split}
   (\tilde\tau_0^\mu)_{\alpha\beta,\gamma\delta} &= \left(\mP_{13}(m_D)\,\mP_{24}(m_{D^\ast})\,\gamma^5_{\alpha\gamma}\,\gamma^\mu_{\beta\delta} - \mP_{13}(m_{D^\ast})\,\mP_{24}(m_{D})\,\gamma^\mu_{\alpha\gamma}\,\gamma^5_{\beta\delta}\right)\,\mC_{11}\,\mF_0\,, \\
   (\tilde\tau_1^\mu)_{\alpha\beta,\gamma\delta} &= \mP_{14}(m_{J/\psi})\,\mP_{23}(m_\omega)\,\varepsilon^{\mu\nu\rho\sigma}\,\hat{P}^\nu\,\gamma^\rho_{\alpha\delta}\,\gamma^\sigma_{\beta\gamma}\,\mC'_{11}\,\mF_0\,, \\
   (\tilde\tau_2^\mu)_{\alpha\beta,\gamma\delta} &= \Big( \mP_{12}(m_S)\,\mP_{34}(m_A)\,(\gamma_5 C)_{\alpha\beta}\,(C^T \gamma^\mu)_{\gamma\delta} \\[-1mm]
                                           & \; \,- \mP_{12}(m_A)\,\mP_{34}(m_S)\,(\gamma^\mu C)_{\alpha\beta}\,(C^T \gamma_5)_{\gamma\delta}\Big) \,\mC_{\bar{3}3}\,\mF_0\,. \\
\end{split}
\end{equation}
Here,
\begin{equation}
    \mP_{ab}(m) = \frac{1}{(p_a+p_b)^2 + m^2}
\end{equation}
is the respective pole contribution, where $p_a$ and $p_b$ are the momenta of the quarks $a$ and $b$. The masses of the mesons ($m_D$, $m_{D^\ast}$, $m_{J/\psi}$, $m_{\omega}$) and the scalar and axialvector heavy-light diquarks ($m_S$, $m_A$)
are calculated beforehand from their two-body BSEs.
Because the poles capture the dominant momentum dependence in the doublet variables,
the resulting four-body amplitude for the $\chi_{c1}(3872)$ can then be written as
\begin{equation}\label{X3872-approx}
   \Gamma^{\mu}(p,q,k,P)_{\alpha\beta\gamma\delta} \approx \sum_{i} f_i(\mS_0)\, (\tilde\tau_i^\mu)_{\alpha\beta,\gamma\delta}\,.
\end{equation}
In this way, the BSE turns into a coupled system of equations for the components $f_i(\mS_0)$, so that the main remaining complication
is the explicit calculation of the four-momentum loop integrals that appear in Fig.~\ref{fig:4q-bse}.

\section{Analytic continuation and extrapolation}

There is one more obstacle to consider when solving the four-body equation in practice.
If the resulting four-quark state is above any of the meson-meson thresholds,
the poles in Fig.~\ref{fig:4q-poles} move inside the integration region and the state becomes a resonance.
However, for a resonance the pole condition $P^2=-M^2$ only has solutions in the complex plane
on a higher Riemann sheet.

To this end, we note that after discretizing the momentum variables the four-quark BSE turns into an eigenvalue equation of the form
\begin{equation}\label{eq:fq-bse-expl}
    \lambda^{(n)}(M)\,f_i^{(n)}(M) = K_{ij}(M)\,f_j^{(n)}(M)\,,
\end{equation}
where $K(M)$ is the kernel matrix and the $f^{(n)}(M)$ are the eigenvectors corresponding to the eigenvalues $\lambda^{(n)}(M)$.
The index $i$ now absorbs both the dependence on the momentum variables and the different amplitudes in Eq.~\eqref{4q-amp}.
$M$ is the variable related to $P^2=-M^2$ and can be real or complex, so the equation holds for bound states and resonances alike.
It has physical solutions whenever  the condition $\lambda^{(n)}(M) = 1$ is satisfied.
The resulting  $M_n$ are the masses of the states,
with $n=0$ for the ground state and $n\geq 1$ for the excitations.

              \begin{figure*}[t]
                    \centering
                    \includegraphics[width=1\textwidth]{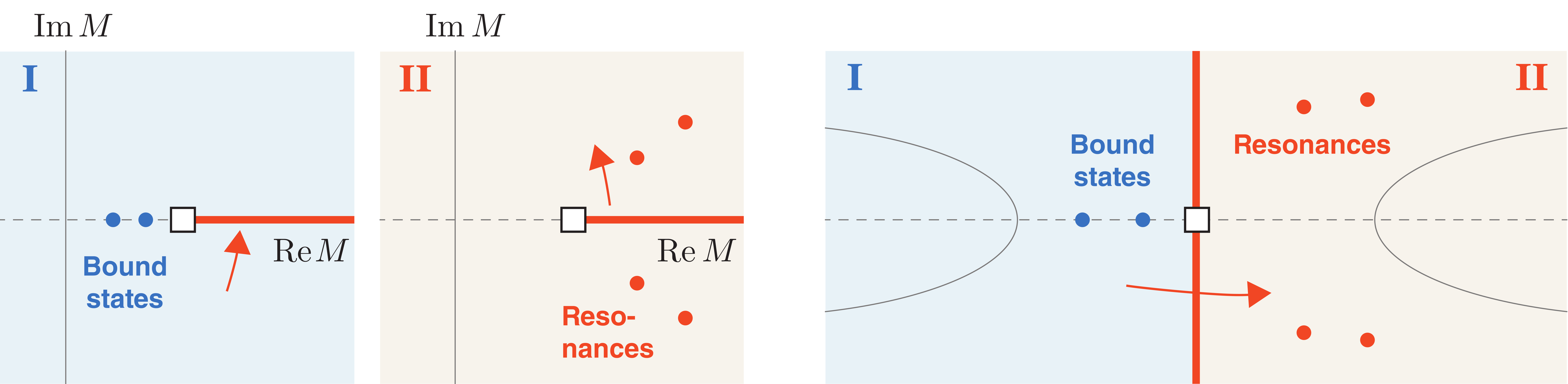}
                    \caption{Sketch of the analytic structure of the BSE eigenvalues, where bound states appear on the first sheet and resonances on the second.
                             Each state corresponds to the condition $\lambda^{(n)}(M) = 1$.
                             Above the branch point (denoted by the square), going from the first to the second sheet is an analytic operation,
                             which can be made explicit by unfolding the branch cut (right diagram).
                     }\label{fig:cuts-2}
            \end{figure*}  

To extract  bound states, it is therefore sufficient to solve the BSE for real $M$  below the first threshold.
To extract resonance locations, one can solve the BSE for complex $M$ on the first sheet above the threshold
and employ analytic continuations to the second sheet, as sketched in Fig.~\ref{fig:cuts-2}. 
However, this requires contour deformations which complicate matters in practice~\cite{Williams:2018adr,Eichmann:2019dts,Santowsky:2020pwd,Santowsky:2021ugd,Miramontes:2021xgn}.
Alternatively, by unfolding the two sheets it is also possible to directly extrapolate from the real axis (right panel in Fig.~\ref{fig:cuts-2}).
Explicit examples for the determination of the $\rho$- and $\sigma$-meson poles using this procedure can be found in Ref.~\cite{Santowsky:2020pwd,Santowsky:2021ugd}.
The situation for heavy-light four-quark states is further complicated by the fact that there are several close-by thresholds in the system.
Therefore, an extrapolation from the real axis only serves as a rough estimate for the true pole position.

To obtain the masses and corresponding errors, % for the four-quark states, 
we apply a two-step procedure which consists of the 
extrapolation of the eigenvalues and the investigation of  $M_n$
with varying current-quark masses.
To  determine $M_n$,
we calculate the eigenvalue in 
Eq.~\eqref{eq:fq-bse-expl} for different masses 
up to the lowest-lying threshold, e.g., $M_{\mathrm{max}}=m_{J/\psi} + m_{\omega}$ for the  $\chi_{c1}(3872)$.
If the resulting eigenvalue curve satisfies the condition $\lambda(M)=1$ in this range, 
we have found a 
bound state; otherwise, we need an analytic continuation to get an estimate of the pole position. % or the real part of the pole mass.

\begin{figure*}[t]
\centering
\includegraphics[width=0.98\textwidth]{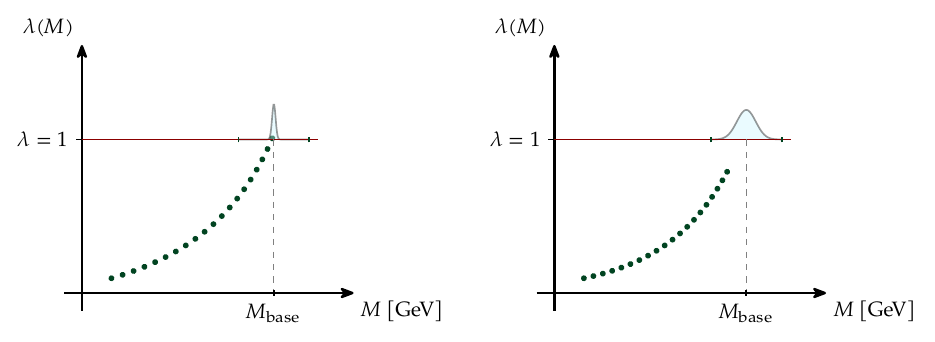}
\caption{\label{fig:evc-extrapolation}Distribution of masses for two different 
eigenvalue curves. $M_{\mathrm{base}}$ is the base mass estimate from the 
polynomial fit and the small vertical lines indicate the chosen acceptance range. \textit{Left:} The eigenvalue curve barely reaches the 
$\lambda(M)=1$ condition and the  distribution of masses is very narrow around 
the intersection. \textit{Right:} Here the extrapolation range is larger and thus also 
the spread of the obtained masses. }
\end{figure*}

To obtain the best statistical estimate, 
we  apply the 
following extrapolation technique. 
Say we have calculated a set of eigenvalues $\lambda_1(M), \dots, \lambda_N(M)$ corresponding to a given state $n$.
We determine a first mass estimate 
$M_{\mathrm{base}}$ by fitting a cubic or quartic polynomial 
to these eigenvalues and extrapolate to $\lambda(M)=1$. 
We then use the Schlessinger-Point Method~\cite{Schlessinger1968}, which is especially good at identifying poles, to analytically continue the 
combinations $1/(1-\lambda_i(M))$ a few hundred times. Because the method is  known to produce spurious singularities,
we keep  those masses within a $5 \dots 10\%$ acceptance region around the base mass.
To improve these estimates, we generate 
random subsets of eigenvalues and repeat the procedure again a few hundred times for each subset. 
The resulting masses in each subset form normal distributions as shown in Fig.~\ref{fig:evc-extrapolation}. 
If the eigenvalue curve satisfies  $\lambda(M)=1$ 
directly (left panel), they are very narrowly centered 
around the base mass, whereas if the extrapolation range is 
larger (right panel)  the spread is  broader.
From these we determine the final mass  $M_n\pm \Delta M_n^{\mathrm{extr.}}$ together with its error estimate.

\begin{figure*}
\centering
\includegraphics[width=1.0\textwidth]{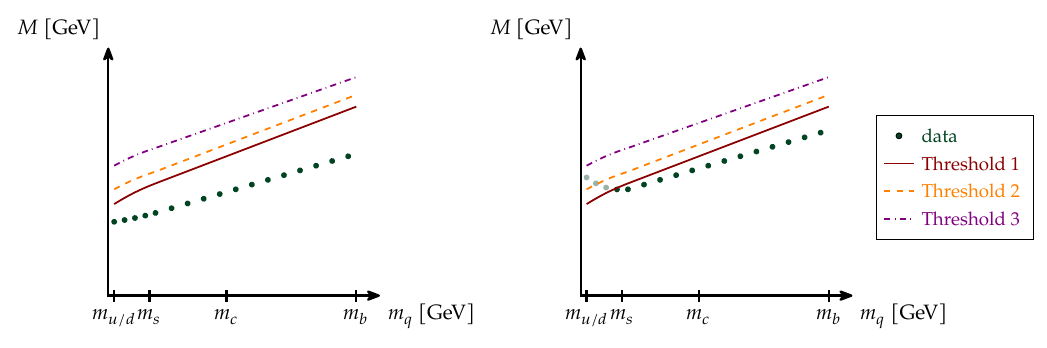}
\caption{\label{fig:qmec} Current-quark mass evolution  for a 
deeply bound  four-quark state
(\textit{left}) and for the case where the four-quark state with light quarks 
is close to or above the lowest threshold (\textit{right}). 
The green points are the calculated data points and the full, dashed and dash-dotted lines are the 
thresholds.}
\end{figure*}

In general, it is interesting to investigate how the masses of the four-quark states behave 
when the current-quark masses are varied. To do so, we keep 
the masses of the heavy quarks ($Q$) fixed 
and vary the current-quark masses of the `light' quarks ($q$) from light to bottom. 
For each current-quark mass, we  apply the procedure described above.
The resulting masses can then be plotted against the `light' current-quark mass, as 
shown in the two examples in Fig.~\ref{fig:qmec}. Here the three curves  represent 
the quark-mass 
evolution of the  two-body thresholds in the system. 
The left panel shows a state that is deeply bound for all quark masses, so  its mass can be determined directly like in the left panel in 
Fig.~\ref{fig:evc-extrapolation}. 
In the right panel, the mass approaches the lowest threshold in the system for decreasing current quark masses and eventually crosses the threshold. 
For the opaque data points the eigenvalue 
curve needs to be extrapolated (right panel in Fig.~\ref{fig:evc-extrapolation}) and typically has  large error bars. 
We now fit the  quark mass evolution curve with the form
\begin{align}\label{eq:fit}
	M(m_q) = \sqrt{\sum_{p=0}^4 c_p\, m_q^p} \,.
\end{align}
%where we apply the fit to the data points below 
%the lowest threshold and obtain the remaining masses via the fit. 
For the cases  shown in the right panel of Fig.~\ref{fig:qmec},  one often observes an upward bending of the 
quark mass evolution curve, which we consider unphysical  
and usually neglect these points when applying the fit.
For a deeply bound state the fit is not strictly necessary since the mass can be anyhow determined directly from the eigenvalue curve. 
The total error is then a combination of the extrapolation error and the 
error coming from the fit of the data. %, i.e, $\Delta M_n = \Delta M_n^{\mathrm{extr.}} + \Delta M_n^{\mathrm{fit}}$.

\section{Results and discussion}\label{sec:results}

\begin{figure*}[t]
                    \centering
                    \includegraphics[width=0.49\textwidth]{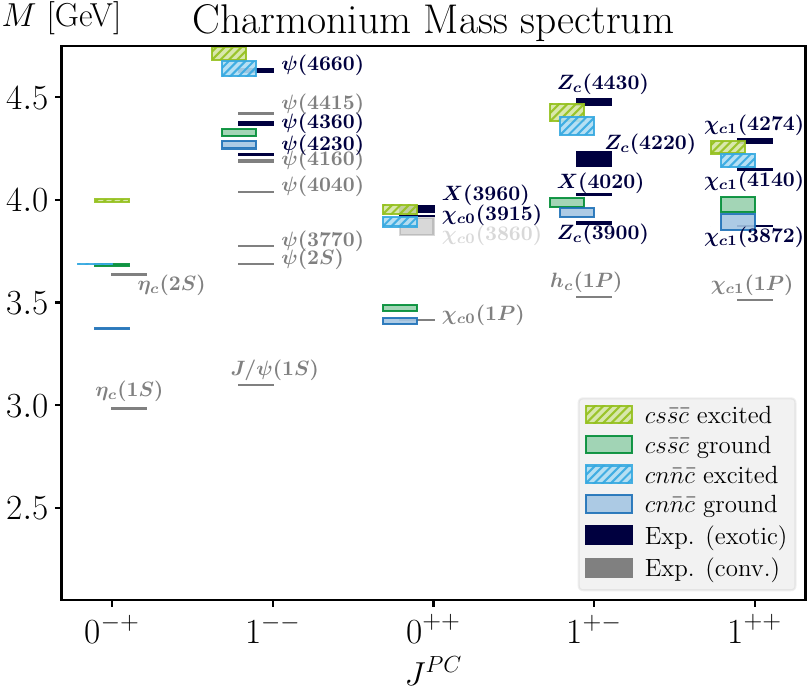}
                    \includegraphics[width=0.49\textwidth]{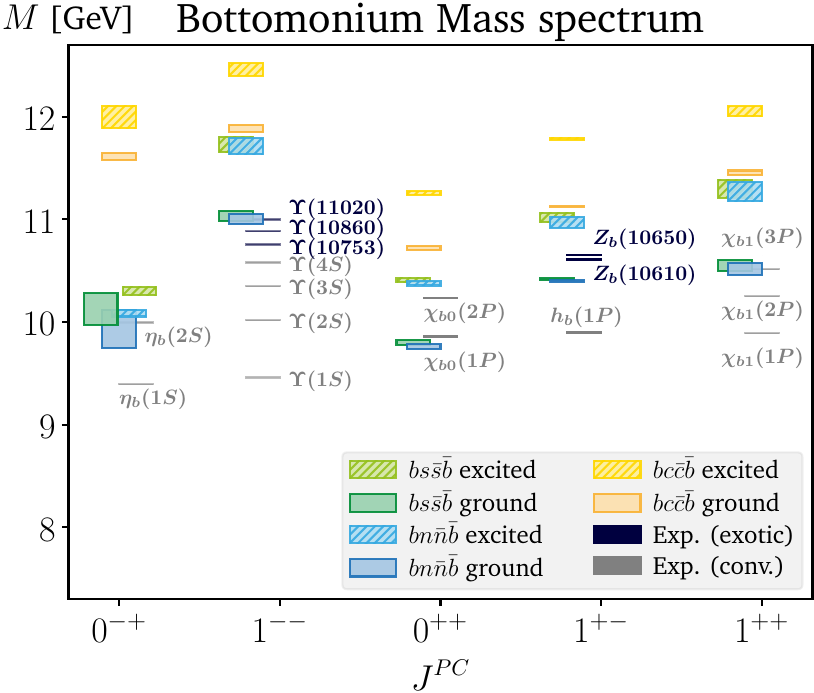}
                    \caption{Charmonium and bottomonium spectrum, with conventional $c\bar{c}$ states from the PDG shown in gray
                             and exotic states in black. The colored boxes are the results from the four-quark equation~\cite{Hoffer:2024alv}.}\label{fig:4q-hidden}
                    \vspace{5mm}
            \end{figure*}

In this section we discuss the results for hidden-flavor states~\cite{Hoffer:2024alv} and open-flavor states~\cite{Hoffer:2024fgm} obtained from the four-quark BSE.
The hidden-flavor charmonium and bottomonium spectra are shown in Fig.~\ref{fig:4q-hidden}. 
The gray levels are the conventional $c\bar{c}$ states from experiment,
the black levels are the exotic candidates, and the colored boxes are the BSE results for $cn\bar{n}\bar{c}$ and $cs\bar{s}\bar{c}$ states 
(ground state and first excited state each). 
One can see that the proximity of the $cn\bar{n}\bar{c}$ and $cs\bar{s}\bar{c}$ states
leads to a densely populated spectrum. 
The $\chi_{c1}(3872)$ in the $1^{++}$ channel is well reproduced.
Note, however, that a backcoupling of the conventional $c\bar{c}$
components is not yet included in this calculation, while lattice calculations indicate that this component might be important~\cite{Prelovsek2013,Padmanath2015}.
An identification for the $1^{+-}$ states is also possible, and
the same is true for the $1^{--}$ vector channel.
An exception is the $0^{++}$ channel, where the four-body results are close to the 
ordinary $c\bar{c}$ state. Similar features are observed in the bottomonium spectrum. 

Concerning the internal structure of these states, the $1^{--}$, $0^{++}$ and $1^{++}$ channels are strongly dominated
by meson-meson configurations (like $D\bar{D}^\ast$ for the $\chi_{c1}(3872)$), while in the $1^{+-}$ channel they are mixtures 
of meson-meson and hadrocharmonium components, where the latter is dominant. The diquark components are almost negligible in all cases.

\begin{figure*}[t]
                    \centering
                    \includegraphics[width=0.67\textwidth]{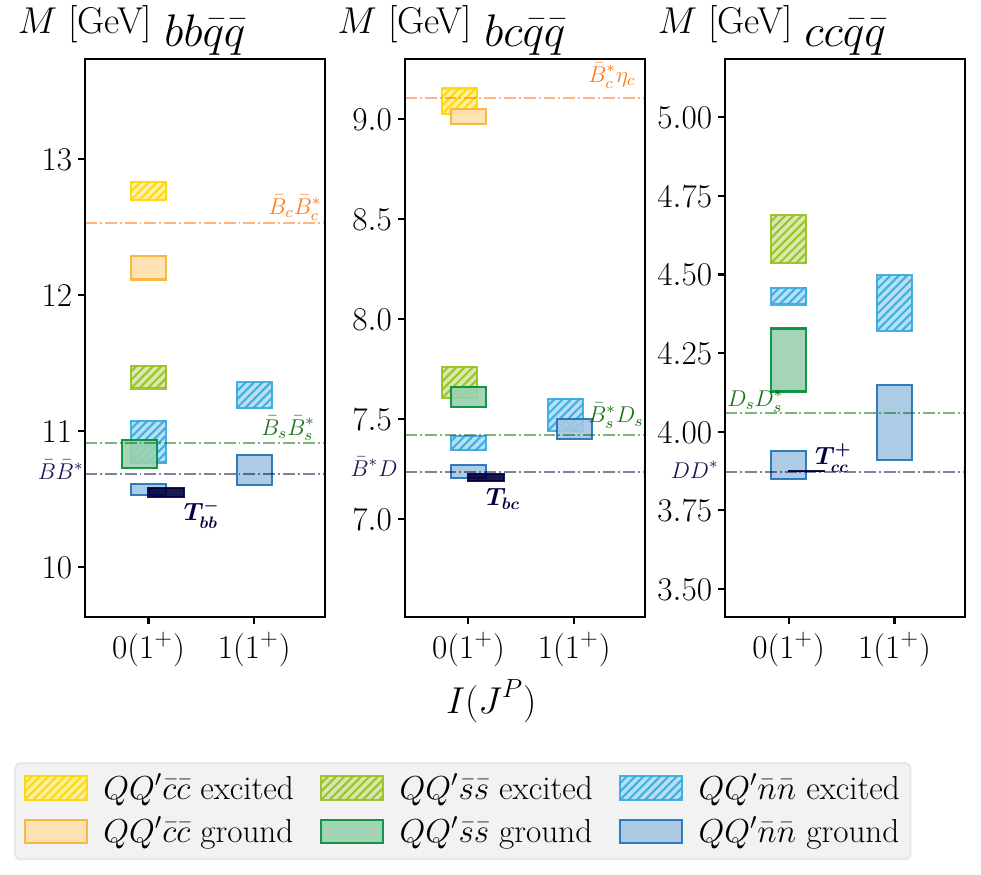}
                    \caption{Open-flavor spectrum for $J^P = 1^+$ and quark content $bb\bar{q}\bar{q}$, $bc\bar{q}\bar{q}$ and $cc\bar{q}\bar{q}$.
                             The levels in black are the experimental $T_{cc}^+$ mass and the theory predictions for its partners.
                             The colored boxes are the four-quark results~\cite{Hoffer:2024fgm}.}\label{fig:4q-open}
            \end{figure*}

For open-charm states the situation is different. Here the two-body clusters are diquark-antidiquark $(QQ)(\bar{q}\bar{q})$
and two degenerate meson-meson configurations $(Q\bar{q})(Q\bar{q})$. 
The physical components taken into account are listed in Table~\ref{tab:components}:
Here, $f_0$ and $f_1$ are meson-meson configurations, $f_2$ is a diquark-antidiquark configuration,
and $f_3$, $f_4$ are further repulsive color octet-octet components which turn out to be important as well.
The resulting spectra for 
$bb\bar{q}\bar{q}$, $bc\bar{q}\bar{q}$ and $cc\bar{q}\bar{q}$ states with $J^P = 1^+$  are shown in Fig.~\ref{fig:4q-open}.
In the rightmost plot, the black level is the experimental mass of the $T_{cc}^+$, while its partners in the other two plots are theory predictions~\cite{Hudspith2023,Alexandrou2024,Junnarkar2019,Leskovec2019,Aoki2023,
      Park2019,Noh2021,Maiani2019,Braaten2021,Alexandrou2024a,Francis2019,Padmanath2024}.
Also here, the BSE results nicely match the ground states.

\begin{table}[!b]
\centering
\begin{tabular}{l||cc|c|cc}
	 $0(1^+)$          & $f_0$          & $f_1$           & $f_2$          & $f_3$     & $f_4$       \\  \hline
	$bb\bar{s}\bar{s}$ & $B_s B_s^\ast$ &                 & $A_{bb}A_{ss}$ & $\bullet$ &             \\
	$bb\bar{n}\bar{n}$ & $B B^\ast$     & $B^\ast B^\ast$ & $A_{bb}S_{nn}$ & $\bullet$ & $\bullet$   \\
	$bc\bar{s}\bar{s}$ & $B_s D_s^\ast$ & $B_s^\ast D_s$  & $S_{bc}A_{ss}$ & $\bullet$ & $\bullet$  \\
	$bc\bar{n}\bar{n}$ & $B D^\ast$     & $B^\ast D$      & $A_{bc}S_{nn}$ & $\bullet$ & $\bullet$  \\
	$cc\bar{s}\bar{s}$ & $D_s D_s^\ast$ &                 & $A_{cc}A_{ss}$ & $\bullet$ &             \\
	$cc\bar{n}\bar{n}$ & $D D^\ast$     & $D^\ast D^\ast$ & $A_{cc}S_{nn}$ & $\bullet$ & $\bullet$  
\end{tabular}\qquad
\begin{tabular}{l||cc|c|cc}
	 $1(1^+)$          & $f_0$      & $f_1$      & $f_2$          & $f_3$     & $f_4$   \\  \hline
	$bb\bar{s}\bar{s}$ &            &            &                &           &  \\
	$bb\bar{n}\bar{n}$ & $B B^\ast$ &            & $A_{bb}A_{nn}$ & $\bullet$ &  \\
	$bc\bar{s}\bar{s}$ &            &            &                &           &  \\
	$bc\bar{n}\bar{n}$ & $B D^\ast$ & $B^\ast D$ & $S_{bc}A_{nn}$ & $\bullet$ & $\bullet$ \\
	$cc\bar{s}\bar{s}$ &            &            &                &           &  \\
	$cc\bar{n}\bar{n}$ & $D D^\ast$ &            & $A_{cc}A_{nn}$ & $\bullet$ & 
\end{tabular}
\caption{\label{tab:components} Physical components of the four-quark amplitudes with $I(J^P) = 0(1^+)$ and $1(1^+)$, 
where $n \in \{u,d\}$ stands for light quarks~\cite{Hoffer:2024fgm}. Scalar and axialvector diquarks are denoted by $S$ and $A$, respectively. 
$f_0$ and $f_1$ are meson-meson configurations $(\mathbf{1}\otimes\mathbf{1})$, $f_2$ is a diquark-antidiquark configuration $(\overline{\mathbf{3}}\otimes\mathbf{3})$,
and $f_3$ and $f_4$ correspond to repulsive octet-octet components $(\mathbf{8}\otimes\mathbf{8})$.
States with isospin $I=1$ cannot exist without light quarks, and the remaining
empty slots are not populated for physics reasons. The repulsive diquark-antidiquark 
configuration $(\mathbf{6}\otimes\overline{\mathbf{6}})$ was also taken into account in 
Ref.~\cite{Hoffer:2024fgm} but turns out to be  negligible.}
\end{table}

\begin{figure*}[t]
                    \centering
                    \includegraphics[width=0.9\textwidth]{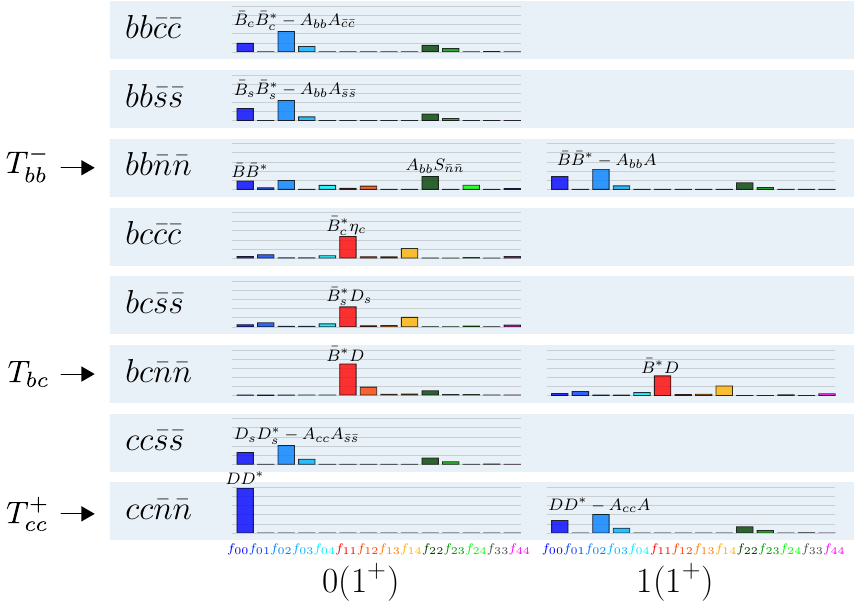}
                    \caption{Leading norm contributions for open-flavor states~\cite{Hoffer:2024fgm}, cf.~Table~\ref{tab:components}.}\label{fig:4q-open-norm}
            \end{figure*}

The internal structure of the open-flavor states is particularly interesting, as shown in Fig.~\ref{fig:4q-open-norm}.
The bars are the norm contributions for each state, which result from the overlap of different wave-function components
with four dressed quark propagators, in shorthand notation:
$f_{ij} = \langle f_i | f_j \rangle$.
The $T_{cc}^+$ in the bottom line of Fig.~\ref{fig:4q-open-norm} 
is almost entirely dominated by the molecular $D D^{\ast}$ component ($f_{00}$) due to its proximity to the threshold. 
Its bottom partner $T_{bb}^-$, on the other hand, is a broad mixture with a dominant diquark-antidiquark component ($f_{22}$).
A similar composition applies for other states containing two bottom quarks, although in those cases the mixing of $f_0$ and $f_2$ is more pronounced.
For the $T_{bc}$, on the other hand, the $B^\ast D$ component is the leading one, and similar conclusions apply
for the other $bc$ states.

One can see that the key strength of the four-body equation lies in its ability to predict the dominant two-body clusters dynamically. 
This contrasts with model-based approaches, where one assumes a specific component (diquark-antidiquark, meson molecule, or hadroquarkonium)
at the outset and makes predictions within that framework. However, the relative importance of these components can vary significantly from case to case or even form a complex admixture, and  therefore it is a dynamical question which structure dominates. 
In the four-body BSE, these components emerge naturally from the underlying quark and gluon interactions without the need for prior assumptions.

%\section{Outlook}

The work discussed here can be extended in several directions. For open-flavor states, it will be valuable to explore 
quantum numbers beyond $J^P = 1^+$. Additionally, states with different quark compositions are of interest, 
such as those involving three different flavors which have been observed at LHCb~\cite{Aaij2021} and BESIII~\cite{Ablikim2021}.
On the technical side, a systematic expansion of the tensor bases beyond the leading physical components is desirable 
in view of aiming towards complete bases, 
as well as an inclusion of the remaining dynamical variables beyond the pole contributions.
In the longer term, progress beyond the rainbow-ladder approximation would be beneficial,
including the incorporation of irreducible three- and four-body interactions in the kernel.

Finally, the results presented here indicate that the approach is sufficiently advanced  to establish a starting point for considering also higher multiquark systems.
Possible applications are the pentaquarks observed at LHCb or also six-quark states, which encode information on the binding 
of nuclei and hypernuclei. Efforts in these directions are underway~\cite{Eichmann:2025gyz,Eichmann:2025etg}.

\newpage

\bigskip
\noindent
\textbf{Acknowledgments:} This work was supported by the Austrian Science Fund FWF under grant number 10.55776/PAT2089624, 
by the BMBF under project number 05P2021, 
the DFG under grant number FI 970/11-2, the graduate school HGS-HIRe, and the GSI Helmholtzzentrum f\"ur Schwerionenforschung.
This work contributes to the aims of the USDOE ExoHad Topical Collaboration, contract DE-SC0023598.

\bibliographystyle{utphys-mod3}
\bibliography{reference}

%\begin{thebibliography}{99}
%\bibitem{...}
%....
%
%\end{thebibliography}

\end{document}